\documentclass[reprint,amsmath,amssymb,aps,pre,floatfix]{revtex4-1}
\usepackage{graphicx}
\usepackage{dcolumn}
\usepackage{siunitx}
\usepackage{dsfont,mathtools}
\usepackage{multirow}
\usepackage[colorlinks=true]{hyperref}

\DeclarePairedDelimiter{\abs}{\lvert}{\rvert}
\DeclareMathOperator{\erfc}{erfc}
\DeclareMathOperator{\supp}{supp}
\newcommand{\df}{\mathrm{d}}
\newcommand{\vv}[1]{\mathbf{#1}}
\newcommand{\te}[1]{\text{#1}}
\newcommand{\ie}{\emph{i.e.}}
\newcommand{\eg}{\emph{e.g.}}
\newcommand{\etal}{\emph{et al.}}

\begin{document}

\title{Grain-scale modeling and splash parametrization\texorpdfstring{\\}{ }for aeolian sand transport}

\author{Marc L\"ammel}%
\author{Kamil Dzikowski}%
\author{Klaus Kroy}%
\email{klaus.kroy@uni-leipzig.de}%
\affiliation{Institut f\"ur Theoretische Physik, Universit\"at Leipzig, Postfach 100920, 04009 Leipzig, Germany }%

\author{Luc Oger}%
\email{luc.oger@univ-rennes1.fr}%
\author{Alexandre Valance}%
\email{alexandre.valance@univ-rennes1.fr}%
\affiliation{Institut de Physique de Rennes, CNRS UMR 6251, Universit\'e de Rennes I, 35042 Rennes, France}%

\date{\today}

\begin{abstract}
  The collision of a spherical grain with a granular bed is commonly parametrized by the splash function, which provides the velocity of the rebounding grain and the velocity distribution and number of ejected grains.
  Starting from elementary geometric considerations and physical principles, like momentum conservation and energy dissipation in inelastic pair collisions, we derive a rebound parametrization for the collision of a spherical grain with a granular bed.
  Combined with a recently proposed energy-splitting model~[Ho~\etal, Phys.~Rev.~E {\bf 85}, 052301 (2012)] that predicts how the impact energy is distributed among the bed grains, this yields a coarse-grained but complete characterization of the splash as a function of the impact velocity and the impactor--bed grain-size ratio.
  The predicted mean values of the rebound angle, total and vertical restitution, ejection speed, and number of ejected grains are in excellent agreement with experimental literature data and with our own discrete-element computer simulations.
  We extract a set of analytical asymptotic relations for shallow impact geometries, which can readily be used in coarse-grained analytical modeling or computer simulations of geophysical particle-laden flows.
\end{abstract}

\maketitle

\section{\label{sec:introduction}Introduction}
Granular flows are ubiquitous in nature and frequently encountered in everyday life.
Their profound understanding is a necessary prerequisite for designing and improving processing steps in industry as well as for predicting hazards like rockfall, avalanches, or devastating shifting sands.
In particular, the grain hopping excited by strong winds shapes arid regions on Earth or other astronomical bodies, thereby creating a whole hierarchy of structures that span orders of magnitude in size.
The collisions of the hopping grains with the sand bed result in a dissipative rebound and grain splashing.
These are essential features that need to be understood to predict aeolian transport and the whole ensuing structure formation.

Since Bagnold's~\cite{Bagnold1941} pioneering investigations back in the 1940s, grain--bed collisions have been studied in wind tunnels~\cite{Willetts1986,Willetts1989,McEwan1992,Rice1995,Rice1996,Dong2002}, by shooting steel or plastic beads onto a quiescent granular bed~\cite{Mitha1986,Oger2005,Crassous2007,Beladjine2007,Oger2008,Ammi2009,Valance2009}, and in (event-driven) computer simulations~\cite{Werner1988,Anderson1988,Anderson1991,Rioual2003,Bourrier2008}.
The general aim is to parametrize the complex stochastic process by the so-called \emph{splash function}~\cite{Ungar1987}.
It provides the average velocity of the rebounding grain after the collision and the average number and the velocity distribution of bed particles ejected in the splash, given the velocity of the impacting particle and the size ratio of the impacting particle to the bed particles.
Unequal grain sizes are of interest, because field observations indicate that aeolian structure formation may be linked to grains sorting, as observed in megaripples~\cite{Yizhaq2009,Yizhaq2012a,Qian2012}.
For their theoretical understanding, a robust and reliable parametrization for the splash function of bidisperse granulates could be of great help.
To establish such a parametrization based on physical arguments and mathematical modeling was a major motivation for the study reported below.

We divide the presentation into two parts, according to the two physical processes at work during the splash process, the rebound of the impacting grain and the impact-driven ejection of bed grains.
The first process appears to be both conceptually and technically less complex, as it can, in a reasonable approximation, be reduced to a two-body scattering problem.
This can be analyzed straightforwardly by means of a combination of elementary geometric considerations and basic physical principles, like momentum conservation.
In the first part of Sec.~\ref{sec:rebound}, we show how various rebound observables, \eg, rebound angles and coefficients of restitution, their dependence on impact angle and impactor--bed grain-size ratio, and their distributions can be obtained from such an approach.
Then, we illustrate that the predictions compare well with experimental data available form the literature and with our own discrete-element computer simulations.
The second process, grain splashing from the bed, is a full-fledged many-body problem that is much harder to grasp and requires a smart ansatz to formalize the complex momentum propagation through the disordered grain packing.
It is analyzed in Sec.~\ref{sec:bed-grain-ejection}, based on a fragmentation model that is applied to the energy-splitting process in the bed, as recently proposed by Ho \etal~\cite{Ho2012}.
The combination of the results from Secs.~\ref{sec:rebound} and \ref{sec:bed-grain-ejection} constitutes a complete and self-contained description of the splash process with various potential applications, as outlined in the concluding section.

\section{\label{sec:rebound}Rebound process}
The starting point of our analysis of the impacting grain's rebound is a purely geometric picture, where the bed packing is approximated by a bumpy wall of infinite mass.
To account for finite-mass effects on the energy dissipation during the rebound, we introduce effective bead--bead restitution coefficients that have the same form but different values as the ``microscopic'' restitution coefficients characterizing binary collisions.
Introducing a phenomelogically motivated dependence on the impactor--bed grain-size ratio allows for a simple and transparent discussion of the geometry that yields manageable analytical predictions.

For pedagogical reasons, we first present a two-dimensional version of our rebound model that can straightforwardly be extended to three dimensions.
As the predictions turn out to be relatively insensitive to the dimensionality, the simpler two-dimensional version suggests itself as the more promising starting point for most practical purposes.
It can easily be solved analytically and provides asymptotic scaling laws for various quantities of interest.

\subsection{\label{sec:2d-collision-model}Two-dimensional collision model}
\begin{figure}
  \centering
  \includegraphics[width=\linewidth]{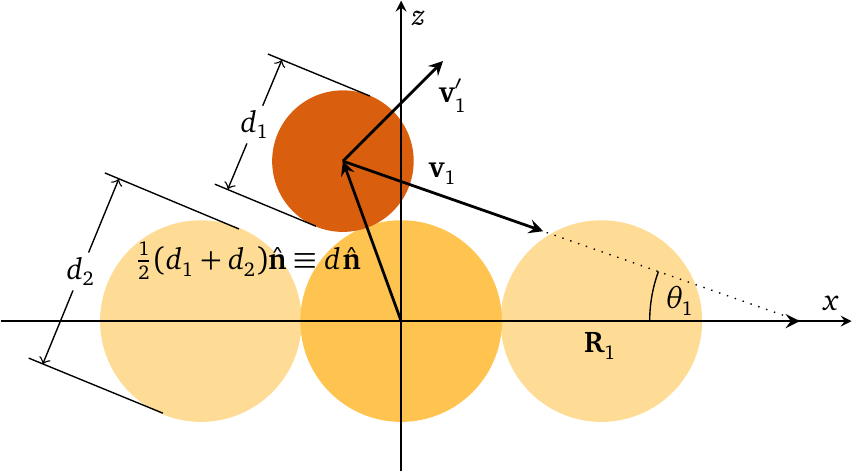}
  \caption{\label{fig:sketch2d}The two-dimensional collision model.
    The spherical impactor of diameter \(d_1\) strikes the homogeneous flat bed with impact velocity \(\vv v_1\) and bounces off the target bed grain of diameter \(d_2\) located at the origin.
    Without the bed grains, the impactor would cross the \(x\) axis at \(\vv R_1\).
    The rebound velocity \(\vv v_1'\) is computed using two independent coefficients of restitution for the tangential and normal component of \(\vv v_1\) according to Eq.~\eqref{eq:restitution-rule}.
    As a consequence, the rebound speed is proportional to the impact speed, \(\abs{\vv v_1'} \propto \abs{\vv v_1}\), which can thus be scaled out, so it suffices to characterize the impactor by its impact angle \(\theta_1\), only.}
\end{figure}
Our formal description of the rebound process is similar to the two-dimensional model proposed by Rumpel~\cite{Rumpel1985}, who considered a regular packing of identical spheres hit by an impactor of the same size as the bed grains.
As a first generalization, we account for different diameters \(d_1\) and \(d_2\) of the impactor and the bed grains, respectively.
In the following, all lengths are given in units of the mean diameter \(d \equiv (d_1 + d_2)/2\), so that \(d_1 + d_2 = 2\).
We denote the impact velocity by \(\vv v_1\) and label all postcollision quantities by a prime.
The collision geometry and the main quantities that we use to describe the rebound are summarized in Fig.~\ref{fig:sketch2d}.
The rebound velocity of the impinging particle, for instance, reads \(\vv v_1'\).
For given impact velocity \(\vv v_1\), the statistical distribution
\begin{equation}
  \label{eq:rebound-distribution}
  \begin{split}
    P(A \vert \vv v_1) &= \overline{ \delta\{ A - f[\vv v_1'(\vv
      v_1,\hat{\vv n})] \} } \\
    &= \frac{1}{d_2} \int_{x_0}^{x_0+d_2}
    \!\!\!\! \!\!\!\! \!\!\!\! \df x \, \delta\{ A - f[\vv v_1'(\vv
    v_1,\hat{\vv n})] \}
  \end{split}
\end{equation}
with mean 
\begin{equation}
  \label{eq:rebound-mean}
  \overline { A}(\vv v_1) = \int \!\! \df A \, A P(A \vert \vv v_1) = \frac{1}{d_2} \int_{x_0}^{x_0+d_2} \!\!\!\! \!\!\!\! \!\!\!\!  \df x \, f[\vv v_1'(\vv v_1,\hat{\vv n})]
\end{equation}
of an observable function \(A = f( \vv v_1')\) of the rebound velocity is obtained by averaging over all possible impact positions \(x\).
We have taken them to be uniformly distributed over the width \(d_2\) of one bed grain.
The implicit \(x\) dependence of \(f\) originates from the normal unit vector \(\hat{\vv n} \equiv \hat{\vv n}(x)\) of the bed surface that is obtained from the collision condition
\begin{equation}
  \label{eq:collision-condition}
  \hat{\vv n} =  \vv R_1 - t \vv v_1
\end{equation}
between the bed grain located at the origin and the impactor that would cross the \(x\)-axis at \(\vv R_1 = (x,0)\) at time \(t=0\), see Fig.~\ref{fig:sketch2d}.
The collision time \(t = \vv R_1 \cdot \vv v_1/\abs{\vv v_1}^2 - (1/\abs{\vv v_1}^2) [(1-\vv R_1^2)\abs{\vv v_1}^2 + (\vv R_1 \cdot \vv v_1)^2]^{1/2}\) is obtained by inserting Eq.~\eqref{eq:collision-condition} into \(\hat{\vv n}^2 = 1\), which yields the normal vector
\begin{equation}
  \label{eq:normal-vector}
    \hat{\vv n} = (\mathds 1 - \hat{\vv v}_1 \hat{\vv v}_1)  \cdot \vv R_1 + \sqrt{1 - \vv R_1 \cdot (\mathds 1 - \hat{\vv v}_1 \hat{\vv v}_1) \cdot \vv R_1 } \; \hat{\vv v}_1
\end{equation}
on the bed-grain surface at the collision point, where \(\hat{\vv v}_1 \equiv \vv v_1/\abs{\vv v_1}\) denotes the normalized velocity vector.
This impact direction is most conveniently characterized in terms of the impact angle \(\theta_1\) as \(\hat{ \vv v}_1 = (\cos \theta_1, \sin \theta_1)\).
Besides the (relative) bed coarseness \(d_2\), the impact angle \(\theta_1\) crucially affects the value of the left-most impact position \(x_0\) in Eqs.~\eqref{eq:rebound-distribution} and \eqref{eq:rebound-mean}.
Determining \(x_0\) from the contact condition between the impactor and the bed grain located at the origin requires us to discriminate between shallow and steep impact angles.
For shallow impact trajectories, the smallest value of \(x_0\) is obtained for a trajectory that is tangential to the left neighbor of the central grain; for steep trajectories, it is determined by the impact position at which the impactor hits the two bed grains at once.
This yields
\begin{equation}
  \label{eq:integration-limits}
   x_0 =
  \begin{dcases}
    \csc \theta_1 - d_2 \,, & 2\sin \theta_1 < d_2 \,, \\
    \cot \theta_1 \sqrt{1 - (d_2/2)^2} - d_2/2 \,, & \te{else.}
  \end{dcases}
\end{equation}
Note that the value of the scaled bed grain diameter \(d_2\) ranges from 0 to 2, corresponding to big and small impactors, respectively.

Before we can evaluate the integrals in Eqs.~\eqref{eq:rebound-distribution} and \eqref{eq:rebound-mean}, we have to specify the (implicit) \(x\) dependence of the function \(f[\vv v_1'(\vv v_1,\hat{\vv n})]\).
To this end, we model the momentum dissipation
\begin{equation}
  \label{eq:restitution-rule}
  \vv v'_1(\vv v_1, \hat{\vv n}) = [- \alpha \, \hat{\vv n} \hat{\vv n} + \beta(\mathds 1 - \hat{\vv n} \hat{\vv n})] \cdot \vv v_1
\end{equation}
during the rebound in terms of the effective restitution coefficients
\begin{equation}
  \label{eq:restitution-coefficients}
  \alpha = \frac{1+\epsilon}{1+\mu} -1 \quad \te{and} \quad \beta = 1-\frac{(2/7)(1-\nu)}{1+\mu} \,, 
\end{equation}
for the normal and the tangential velocity component of the rebounding grain, respectively.
The expressions for \(\alpha\) and \(\beta\) are derived for two inelastically colliding spheres of mass ratio \(\mu\), where the total momentum is taken to be conserved and the energy dissipation is determined by the \emph{microscopic} restitution coefficients \(\epsilon\) and \(\nu\) for the normal and tangential component of the relative surface velocity of the collision partners.
(The latter can be taken to be material parameters, independent of the grain sizes and impact speed.)
We outline this classical calculation in Appendix~\ref{sec:inel-binary-coll}.
To account for the non-trivial grain-size dependence of the energy dissipation in the bed, we freely interpret the grain-mass ratio
as an effective parameter that interpolates between the exactly known asymptotic scaling \(\mu \sim d_1^3/d_2^3\)~\cite{Brilliantov2010} for small grain-size ratios \(d_1 / d_2 \ll \pi/2\) and a phenomenological value that accounts for the high number of excited bed grains for large \(d_1/d_2 \to \infty\).
In the first case, energy transfer to the packing becomes negligible, so the rebound process reduces to a binary collision.
In the second case, for large impactors (\(d_1/d_2 \gg 1\)), the large number of excited bed grains makes the collision highly dissipative, for which we impose the limit \(\mu \sim \epsilon\), so the normal rebound velocity vanishes (\(\alpha \to 0\)).
The expression
\begin{equation}
  \label{eq:effective-massratio}
  \mu = \epsilon d_1^3/\left(d_1^3 + \epsilon d_2^3 \right) \,,
\end{equation}
provides a plausible parametrization that fulfills both of these conditions.

Inserting Eq.~\eqref{eq:normal-vector} for the normal vector \(\hat{\vv n}\) into Eq.~\eqref{eq:restitution-rule} and recalling that \(\vv R_1=(x,0)\), we get the wanted dependence of \(f[\vv v_1'(\vv v_1,\hat{\vv n})]\) on the impact position \(x\).
In particular, for the identity function \(f[\vv v_1'(\vv v_1,\hat{\vv n})] = \vv v_1'(\vv v_1,\hat{\vv n})\), we obtain the components
\begin{subequations}
  \label{eq:2d-rebound-velocity}
  \begin{align}
    \begin{split}
      \frac{v_{1x}'}{\abs{\vv v_1}} &= - \alpha \cos \theta_1 +
      (\alpha+\beta) x^2 \sin^2\theta_1 \cos
      \theta_1 \\
      & \quad + (\alpha + \beta) x \sin^2\theta_1
      \sqrt{1-x^2\sin^2\theta_1}
    \end{split}
    \label{eq:2d-rebound-velocity-x}\\
    \begin{split}
      \frac{v_{1z}'}{\abs{\vv v_1}} &= \alpha \sin\theta_1 - (\alpha+\beta)  x^2 \sin^3\theta_1 \\
      & \qquad + (\alpha+\beta) x \sin\theta_1 \cos\theta_1
      \sqrt{1-x^2\sin^2\theta_1}
    \end{split}
    \label{eq:2d-rebound-velocity-z}
  \end{align}
\end{subequations}
of the rebound velocity \(\vv v_1' = (v_{1x}',v_{1z}')\).
It follows that the rebounding grain continues moving downward into the bed after the first collision if \(1/(1+\beta/\alpha) < x^2 \sin^2\theta_1 - x \cos\theta_1 \sqrt{1- x^2\sin^2\theta_1}\).
In this case, it thus collides with an adjacent second bed grain.
In the remainder of the current section, we neglect such secondary collisions for simplicity, which allows us to derive analytically manageable expressions for the asymptotic scaling of various averages.
Further below, it is demonstrated that only marginal errors are incurred by this approximation.

\subsubsection{\label{sec:2d:shallow-impacts}Shallow impacts \texorpdfstring{(\(\theta_1 \ll \pi/2\))}.}
To facilitate the following analysis, we now make the \(x\) dependence of the function \(f\) in Eq.~\eqref{eq:rebound-distribution} explicit and identify \(f(x) = f[\vv v_1'(\vv v_1,\hat{\vv n})]\).
The \(x\)-integral in Eq.~\eqref{eq:rebound-distribution} can be evaluated if all branches \(f_i^{-1}\) of the inverse of \(f(x)\) are known.
For shallow impacts, \(\theta_1 \ll \pi/2\), there exists only a single branch and the rebound distribution evaluates to
\begin{equation}
  \label{eq:shallow-impacts:rebound-distribution}
  P(A \vert \vv v_1)  \sim
  \begin{dcases}
    \frac{1}{d_2} \abs*{\frac{\df f^{-1}}{\df A}}
 \,, &  0 < \csc \theta_1 - f^{-1}(A) < d_2 \,, \\
    0 \,, & \te{else,}
  \end{dcases}
\end{equation}
where we inserted the first line of Eq.~\eqref{eq:integration-limits} for \(x_0\).
As an example for an interesting observable, consider the rebound angle, \ie, \(f(x) = \theta_1' = \arctan(v_{1z}'/v_{1x}')\).
Replacing \(x\) in Eq.~\eqref{eq:2d-rebound-velocity} by the shifted coordinate \(x + d_2 - \csc \theta_1\), we obtain its exact shallow-impact asymptotics \(\theta_1' \sim (1+\alpha/\beta) \sqrt{2(d_2-x)\theta_1} - \theta_1\).
Inserting this into Eq.~\eqref{eq:shallow-impacts:rebound-distribution}, we calculate the statistical distribution
\begin{equation}
  \label{eq:shallow-impacts:rebound-angle-distribution}
  P(\theta_1' \vert \theta_1) \sim
  \begin{dcases}
    \frac{\beta^2(\theta_1+\theta_1')}{(\alpha+\beta)^2  d_2
      \theta_1} \,, &
    0 < \frac{\beta(\theta_1+\theta_1')}{(\alpha + \beta) \sqrt{2 d_2\theta_1}} < 1 \,, \\
    0 \,, & \te{else,}
  \end{dcases}
\end{equation}
and mean
\begin{equation}
  \label{eq:shallow-impacts:rebound-angle-mean}
  \overline{\theta_1'} \sim (2/3)(1 + \alpha/\beta)\sqrt{2 d_2 \theta_1} - \theta_1 \,.
\end{equation}
of the rebound angle.
The same procedure can be applied to the total and the vertical restitution \(e \equiv \abs{\vv v_1'}/\abs{\vv v_1}\) and \(e_z \equiv v_{1z}'/\abs{v_{1z}}\), respectively.
For small \(\theta_1\), \(e \sim \beta - (\beta^2-\alpha^2)(d_2 - x) \theta_1/\beta\) and \(e_z \sim -\beta + (\alpha+\beta)\sqrt{2(d_2-x)/\theta_1}\) follows from Eq.~\eqref{eq:2d-rebound-velocity} after shifting the \(x\) coordinate by \(\csc \theta_1 - d_2\).
Inserting these asymptotically exact results into Eq.~\eqref{eq:shallow-impacts:rebound-distribution} yields
\begin{align}
  \label{eq:shallow-impacts:total-restitution-distribution}
  P(e\vert \theta_1) &\sim
  \begin{dcases}
    \frac{\beta}{(\beta^2-\alpha^2) d_2 \theta_1} \,, & 0 < \frac{\beta(\beta-e) }{(\beta^2-\alpha^2) d_2 \theta_1} < 1 \,, \\
    0 \,, & \te{else,}
  \end{dcases}\\
  \label{eq:shallow-impacts:total-restitution}
  \overline e &\sim \beta - (\beta^2-\alpha^2) d_2 \theta_1/(2\beta) \,,
\end{align}
and 
\begin{align}
  \label{eq:shallow-impacts:vertical-restitution-distribution}
  P(e_z\vert \theta_1) &\sim
  \begin{dcases}
    \frac{(e_z+\beta) \theta_1}{(\alpha+\beta)^2 d_2} \,, & 0 < \frac{ e_z+\beta }{ (\alpha+\beta)\sqrt{2 d_2/\theta_1} } < 1 \,, \\
    0 \,, & \te{else,}
  \end{dcases} \\
  \label{eq:shallow-impacts:vertical-restitution}
  \overline {e_z} &\sim -\beta + (2/3) (\alpha + \beta) \sqrt{2 d_2/\theta_1} \,.
\end{align}

\subsubsection{\label{sec:2d:steep-impacts}Steep impacts \texorpdfstring{(\(\theta_1 \approx \pi/2\))}{}.}
For steep impacts, similar relations can be derived by expanding the observables \(\theta_1'\), \(e\), and \(e_z\), introduced above, in the impact angle \(\theta_1\) up to linear order around \(\pi/2\).
This strategy provides analytical expressions for their mean values, obtained by averaging over the impact position \(x\), but gives no access to their distributions.
The reason is that the \(x\)-dependence \(A=f(x)\) of an observable \(A\) is now nonlinear (in contrast to the shallow-impact expansions), which precludes the inversion \(x=f^{-1}(A)\), required in Eq.~\eqref{eq:rebound-distribution}.
Reasonable approximations of the distributions can nevertheless be obtained by expanding \(f(x)\) up to first or second order in \(x\), as documented in Appendix~\ref{sec:2d-steep-impacts} together with further technical details of the steep-impact expansion.
Here, we only quote the first-order asymptotics of the mean rebound angle, the mean total restitution, and the mean vertical restitution, respectively:
\begin{align}
  \label{eq:steep-impacts:rebound-angle-mean}
  \begin{split}
    \overline{\theta_1'} &\sim  \pi- \theta_1  + \sin^{-1}( d_2/2 ) \sqrt{4/d_2^2-1} (\theta_1-\pi/2)\\
    & \quad + \tan^{-1}\!\left( \frac{ \beta /\alpha}{\sqrt{4/d_2^2-1}} \right) \sqrt{4/d_2^2-1}  (\theta_1-\pi/2) \,,
  \end{split} \\
  \label{eq:steep-impacts:total-restitution-mean}
  \begin{split}
    \overline e &\sim (1/4)\sqrt{4 \alpha ^2 + (\beta^2 -\alpha ^2)
      d_2^2} \\
    & \quad + \frac{\alpha^2}{(\beta^2-\alpha^2)d_2}
    \tanh^{-1}\!\left(\frac{\beta^2-\alpha^2}{4 \alpha^2 +
        (\beta^2-\alpha^2) d_2^2}\right)\,,
  \end{split} \\
  \label{eq:steep-impacts:vertical-restitution-mean}
  \overline{ e_z} &\sim \alpha - (\alpha+\beta) d_2^2/12\,.
\end{align}

\subsubsection{\label{sec:2d:full-solution}Full numerical solution.}
\begin{figure}
  \centering
  \includegraphics[width=\linewidth]{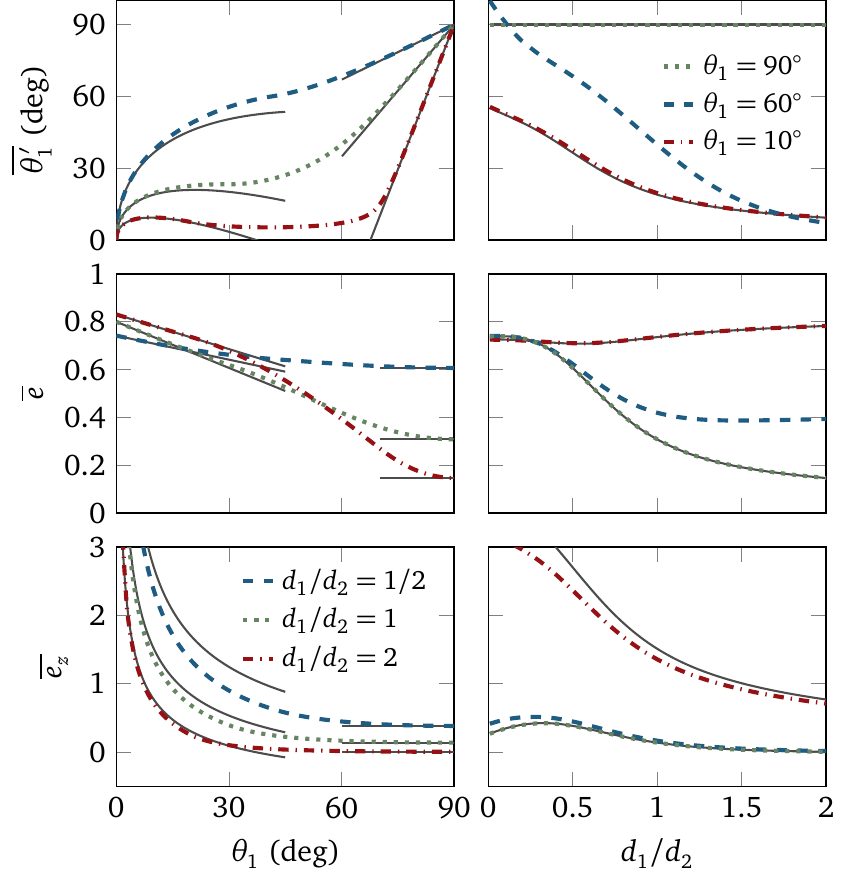}
  \caption{\label{fig:2dng-theez-means}The two-dimensional collision model predicts the mean rebound angle \(\theta_1' = \arctan(v_{1z}'/v_{1x}')\), the total restitution \(e \equiv \abs{\vv v_1'}/\abs{\vv v_1}\), and the vertical restitution and \(e_z \equiv v_{1z}'/\abs{v_{1z}}\) as a function of the impact angle \(\theta_1\) and the impactor--bed grain-size ratio \(d_1/d_2\) from an average over the possible impact positions of the regularly packed bed.    
    Solid lines represent the asymptotic scaling relations, Eqs.~\eqref{eq:shallow-impacts:rebound-angle-mean}, \eqref{eq:shallow-impacts:total-restitution}, \eqref{eq:shallow-impacts:vertical-restitution}, and \eqref{eq:steep-impacts:rebound-angle-mean}--\eqref{eq:steep-impacts:vertical-restitution-mean} derived for shallow and steep impacts, respectively.
    The microscopic restitution coefficients are \(\epsilon=0.75\) and \(\nu=0\).}
\end{figure}

The numerical solution of the proposed rebound model is illustrated in Fig.~\ref{fig:2dng-theez-means}, where \(\overline {\theta_1'}\), \(\overline e\), and \(\overline {e_z}\) are plotted against the impact angle \(\theta_1\) and the impactor--bed grain-size ratio \(d_1/d_2\).
The plots reveal that the asymptotic relations for shallow and steep impacts, Eqs.~\eqref{eq:shallow-impacts:rebound-angle-mean}, \eqref{eq:shallow-impacts:total-restitution}, \eqref{eq:shallow-impacts:vertical-restitution}, and \eqref{eq:steep-impacts:rebound-angle-mean}--\eqref{eq:steep-impacts:vertical-restitution-mean}, shown as dashed lines, indeed provide useful expressions if \(\theta_1 < \SI{20}{\degree}\) and \(\theta_1 > \SI{80}{\degree}\), respectively.
Note that small impactors with a large enough impact angle are scattered backwards (\ie, to the left in the sketch of Fig.~\ref{fig:sketch2d}) for most possible impact positions, yielding a mean rebound angle larger than \SI{90}{\degree}, as it is the case for the \SI{60}{\degree}-impacts shown in the upper right panel of Fig.~\ref{fig:2dng-theez-means} (green curve).

So far, we kept the two-dimensional formulation as simple as possible in order to derive the above analytical relations.
One might suspect this approach to be too simplistic and therefore prone to some unphysical artifacts.
To obtain a more realistic rebound description, we therefore account for a second recoil from the bed grains, thereby suppressing negative (or too large) rebound angles.
This refinement provides only minor quantitative corrections that can approximately be subsumed into a moderate renormalization of the microscopic restitution coefficients \(\epsilon\) and \(\nu\), as illustrated in Sec.~\ref{sec:exp-sim}, where we compare the various model versions with experimental data and our computer simulations.
More significant consequences on the rebound statistics are obtained from a three-dimensional extension of our model that we present in the following section.

\subsection{\label{sec:3d-collision-model}Three-dimensional collision model}
To extend the above toy model to three dimensions, we represent the bed surface by a periodic hexagonal plane packing of spheres.
On contact, the vector connecting the centers of the impacting grain and the bed grain with lattice coordinates \((i,j)\) reads
\begin{equation}
  \label{eq:3d:center-distance}
  \vv R_{1ij} \equiv [ \vv R_1-(i+j/2) d_2 \vv e_x - (\sqrt{3} /2) j d_2 \vv e_y ] \,.
\end{equation}
Replacing its 2D analog \(\vv R_{1i}\) by this expression, all equations of Sec.~\ref{sec:2d-collision-model} stay the same for the 3D model version.
But now, the average is over three parameters: two dimensions of initial position and the horizontal angle of incidence, which by symmetry needs to vary by \(\pi/3\).

To compare with experiments employing a single camera, we have to replace the rebound velocity with its projection
\begin{equation}
  \label{eq:3d:projection}
  \vv v_{1,\te {i}z}' = (\mathds 1-\vv q \vv q) \cdot \vv v_1' \,, \qquad \vv q \equiv \vv v_1 \times \vv e_z / \abs{\vv v_1 \times \vv e_z}\,,
\end{equation}
on the plane viewed by the camera, for which we assume that it is always the plane of incidence (\ie, spanned by \(\vv v_1\) and the \(z\) axis).
As already pointed out above, such a collision can result in a rebound velocity that points downwards, thus leading to secondary collisions.
Altogether, we thus end up with four versions of our geometrical collision model: two or three dimensional, each with one or two bed collisions, which we label in the following as 2D, 2D2, 3D, and 3D2, respectively.
To keep the following analysis and comparison with numerical and experimental data manageable, we will not explicitly consider the 3D model version, because it would neither provide any qualitatively new insight nor is it computationally much more efficient, as we have to evaluate both the 3D and the 3D2 version numerically, anyway.

\begin{figure}
  \centering
  \includegraphics[width=\linewidth]{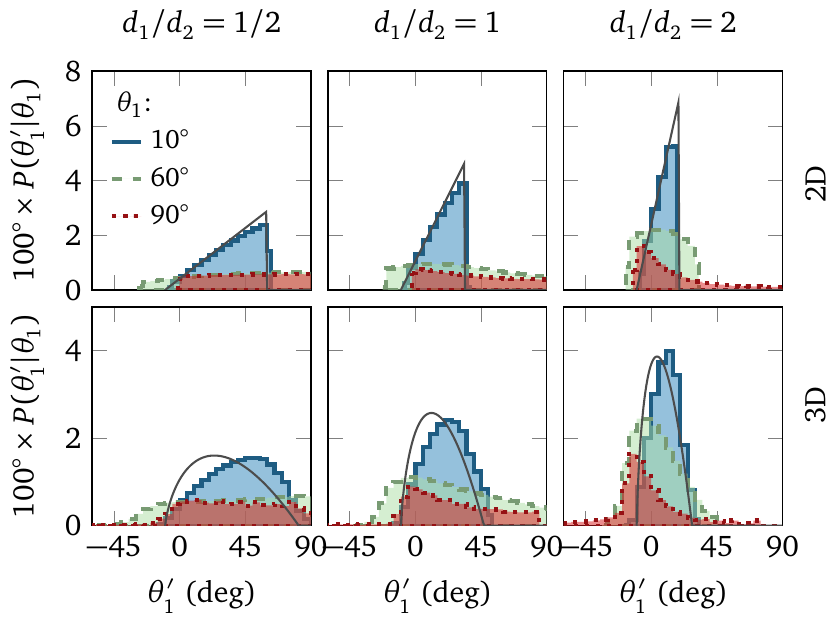}
  \caption{\label{fig:rebound-angle-distribution}The rebound angle distribution \(P(\theta_1' \vert \theta_1)\) as obtained from the 2D and 3D version of our model.
    The qualitative differences are most significant for small impact angles \(\theta_1\) and can be traced back to the irregular packing of two-dimensional slices through a regular three-dimensional granular packing.
    We effectively account for this effect by introducing a uniform distribution of void sizes between neighboring bed grains in the 2D model.
    Thereby, we can analytically determine the asymptotic form of \(P(\theta_1' \vert \theta_1)\) for shallow impacts, Eq.~\eqref{eq:2dvoid:shallow-impacts:rebound-angle-distribution} (dashed lines in the lower panels), which reproduces the bell shape obtained from the numerically evaluated full 3D model rather than the triangular shape predicted by Eq.~\eqref{eq:shallow-impacts:rebound-angle-distribution} (dashed lines in the upper panels).
    The microscopic restitution coefficients are \(\epsilon=0.75\) and \(\nu=0\).}
\end{figure}

To evaluate the difference between the two- and the three-dimensional description, we compare the distribution \(P(\theta_1'\vert \theta_1)\) of the rebound angle \(\theta_1'\) for given impact angle \(\theta_1\) obtained from the 2D and 3D model version for various \(\theta_1\) and \(d_1/d_2\) in Fig.~\ref{fig:rebound-angle-distribution}.
The plot reveals that both approaches yield rather similar distributions, suggesting that the much simpler 2D version suffices in most applications.
Only for shallow impacts, we find that the two- and three-dimensional approaches differ qualitatively:
While the graph of the 2D distribution has a triangular shape, as we also expect form the asymptotic Eq.~\eqref{eq:shallow-impacts:rebound-angle-distribution}, its 3D analog appears to be smoothed, with a considerable contribution of large rebound angles.
The latter is a consequence of the fact that an impactor can reach relatively low---and thus quite steep---parts of a 3D packing when it approaches the trough formed by three neighboring bed grains.
This three-dimensional effect can also be rationalized by considering the 2D slices that cut through the regular three-dimensional packing.
They appear to be highly irregular, characterized by strong variations of the size of the solid disks and intermittent voids.
It should thus be possible to effectively simulate the effects due to the three-dimensional collision geometry by introducing some surface irregularity into the two-dimensional bed.
This idea is addressed in the next section.

\subsection{\label{sec:2d-void}Effective disordered two-dimensional bed}
To effectively simulate the irregularities encountered along two-dimensional cuts though a three-dimensional bed, we introduce a distribution \(p(s_\te d, s_\te v)\) of the reduced disk and void size \(s_\te d\) and \(s_\te v\), respectively.
Here \(s_\te d d_2\) is the actual size of the effective two-dimensional bed grains at the surface and \(s_\te v d_2\) the void size between them.
For each pair \((s_\te d, s_\te v)\), the bed may be considered a regular lattice with periodicity \((s_\te d + s_\te v)d_2\).
The distribution and mean of an observable \(A(\vv v_1)\) that characterizes the rebound of an impactor of velocity \(\vv v_1\) now read
\begin{equation}
  \label{eq:2dvoid:rebound-distribution}
    P(A \vert \vv v_1) =  \int \!\! \df s_\te d \df s_\te v  \frac{p(s_\te d, s_\te v)}{(s_\te d + s_\te v)d_2} \int_{x_0-s_\te v d_2}^{x_0+s_\te d d_2} \hspace{-3em} \df x \, \delta\{ A - f[\vv v_1'(\vv v_1,\hat{\vv n})] \}
\end{equation}
and
\begin{equation}
  \label{eq:2dvoid:rebound-mean}
  \overline { A}(\vv v_1) 
  =  \int \!\! \df s_\te d \df s_\te v \frac{p(s_\te d, s_\te v)}{(s_\te d + s_\te v)d_2} \int_{x_0-s_\te v d_2}^{x_0+s_\te d d_2} \hspace{-3em}
  \df x \, f[\vv v_1'(\vv v_1,\hat{\vv n})] \,,  
\end{equation} respectively.
The integrals in Eqs.~\eqref{eq:2dvoid:rebound-distribution} and \eqref{eq:2dvoid:rebound-mean} average over the void to the left of the bed grain hit by the impactor, whereby the range of the possible impact positions \(x\) is determined by
\begin{equation}
  \label{eq:2dvoid:integration-limits}
   x_0 =
  \begin{dcases}
    \csc \theta_1 - s_\te d d_2 \,, &  \hspace{-0.4\linewidth} 2\sin \theta_1 < (s_\te d + s_\te v)  d_2 \,, \\
    \cot \theta_1 \sqrt{1 - [(s_\te d + s_\te v) d_2/2]^2} - (s_\te d - s_\te v) d_2/2 \,, \\  
    & \hspace{-0.4\linewidth} \te{else.}
  \end{dcases}
\end{equation}
As the \(s\)-dependence of \(x_0\) obviously complicates the calculation of the double integral in Eq.~\eqref{eq:2dvoid:rebound-distribution}, it is useful to bring it into a more transparent form.
Following the formalism introduced in Sec.~\ref{sec:2d-collision-model}, we again make the \(x\)-dependence of the function \(f(x) = f[\vv v_1'(\vv v_1,\hat{\vv n})]\) in Eq.~\eqref{eq:2dvoid:rebound-distribution} explicit.
The \(x\)-integral evaluates to the sum \(\sum_i \abs*{ \df f_i^{-1} / \df A }\) over all branches \(f_i^{-1}\) of the inverse of \(f\).
In general, the support of each \(f_i^{-1}\) is not the whole image \(f([x_0-s_\te v d_2, x_0+s_\te d d_2])\) of \(f\), but only a subset of it.
The integral thus becomes
\begin{equation}
\label{eq:2dvoid:rebound-distribution-s-integral}
  P(A \vert \vv v_1)
  = \sum_i \! \int \!\! \df s_\te d \df s_\te v \frac{p(s_\te d, s_\te v)}{(s_\te d + s_\te v)d_2} \abs*{\frac{\df f_i^{-1}}{\df A}} 
\chi_{i}(A) \,, 
\end{equation}
where the characteristic function \(\chi_{i}(A)\) is one if \(A\) is in the support of \(f^{-1}_i\) and zero otherwise.
Note that \(\chi_{i}(A)\) depends implicitly on \(s\), as \(\supp f_i^{-1}\) is a subset of the \((s_\te v, s_\te d)\)-dependent image \(f([x_0-s_\te v d_2, x_0+ s_\te d d_2])\).

To determine the distribution \(p(s_\te d, s_\te v)\) is again a purely geometrical, though complex, problem, which is further complicated by the fact that, for a two-dimensional cut through a three-dimensional packing,  \(s_\te d\) and \(s_\te v\) are not independent of each other.
Here, we thus propose a rather simplistic approach that can be easily used to derive a number of results.
Testing them against experiments, computer simulations, and (in particular) the full 3D model introduced above, eventually allows us to assess the quality of our simplified model for \(p(s_\te d, s_\te v)\).
First, we neglect the correlations between \(s_\te d\) and \(s_\te v\) and assume that the effect of the bed surface irregularities is sufficiently characterized by a fixed bed grain size \(d_2\) and a uniform void-size distribution:
\begin{equation}
  \label{eq:2dvoid:uniform-void-size-distribution}
  p(s_\te d, s_\te v) \approx \delta(s_\te d - 1) 
  \begin{cases}
    1/(b-a) \,, & a < s_\te v < b \,, \\
    0 \,, &\te{else}.
  \end{cases}
\end{equation}
The complex dependence on the grain-size ratio and the impact angle is thereby delegated to the values of \(a\) and \(b\) that serve to parametrize the effect of the three-dimensional scattering geometry in terms of the two-dimensional void size.
In general, the minimum void size \(b\) can take negative values, but must be larger than \(-1\), as the target grain becomes completely screened by its left neighbor for \(s_\te v = -1\).
The maximum void size \(b\) can basically take arbitrarily large values.

The precise geometry for higher impact angles is somewhat involved.
However, the comparison between the two- and three-dimensional approaches in Fig.~\ref{fig:rebound-angle-distribution} reveals that the rebound statistics differ qualitatively  only for shallow impacts , whereas for steeper impact angles the simple two-dimensional framework already seems to capture the main characteristics of the impact statistics for the three-dimensional geometry quite well.
We thus restrict the following analysis to very shallow impacts.

\subsubsection*{\label{sec:2d-void:shallow-impacts}Shallow impacts \texorpdfstring{(\(\theta_1 \ll \pi/2\))}{}.}
For shallow impacts, the expressions for the minimum and maximum void size take the simple form
\begin{equation}
  \label{eq:2dvoid:shallow-impacts:voidsize-limits}
  a \sim -1 \,, \qquad b \sim -1 + \sqrt 3
  \,,
\end{equation}
respectively, independent of the grain-size ratio and impact angle.
The target bed grain is completely screened for the minimum \(s_\te v =a\), while the impactor is tangent to the next neighbor in front of the trough-forming bed grains for the maximum \(s_\te v = b\).

As the inverse \(f^{-1}\) of \(f(x)\) consists of only a single branch, both \(f\) and \(f^{-1}\) are monotonic functions and the effect of the characteristic function \(\chi_1(A)=1\) is equivalent to the interval condition \(-s_\te v d_2 < f^{-1}(A) < s_\te d d_2\), for which the rebound distribution in Eq.~\eqref{eq:2dvoid:rebound-distribution-s-integral} becomes
\begin{equation}
  \label{eq:2dvoid:rebound-distribution-single-branch}
    P(A \vert \vv v_1) \sim \frac{1}{d_2} \abs*{\frac{\df f^{-1}}{\df
        A}}  \int_{s_0}^\infty \hspace{-1.3em} \df s_\te d
     \int_{-s_0}^\infty \hspace{-1.4em} \df s_\te v \frac{p(s_\te d, s_\te v)}{s_\te d + s_\te v}
\end{equation}
with \(s_0 \equiv f^{-1}(A)/d_2\).
Here we shifted the \(x\) coordinate by \(x_0 = \csc \theta_1 - d_2\), as given in the first line of Eq.~\eqref{eq:2dvoid:integration-limits}.
The latter requires that \(2\sin\theta_1 < (1+s_\te v )d_2\), which is actually not always fulfilled, because \(s_\te v\) becomes as small as \(a=-1\), according to Eq.~\eqref{eq:2dvoid:shallow-impacts:voidsize-limits}.
For small \(\theta_1\), however, we argue that the so introduced error is inconsequential compared to the approximation of the void- and disk-size distributions.
Inserting the uniform void size-distribution, Eq.~\eqref{eq:2dvoid:uniform-void-size-distribution}, with the asymptotic estimates of \(a\) and \(b\) given in Eq.~\eqref{eq:2dvoid:shallow-impacts:voidsize-limits}, we can perform the \(s_\te v\)-integration to obtain the following compact form of the rebound distribution:
\begin{widetext}
  \begin{equation}
    \label{eq:2dvoid:shallow-impacts:rebound-distribution}
    P(A \vert \vv v_1)  \sim  
    \begin{dcases} 
      \frac{\abs*{\df f^{-1}/\df A}}{\sqrt 3 d_2}
      \ln\frac{\sqrt 3 d_2}{d_2-f^{-1}(A)} \,,  & 0 < d_2 - f^{-1}(A) < \sqrt 3 d_2 \,, \\
      0 \,, & \te{else.}
    \end{dcases}
  \end{equation}
  Substituting the shallow-impact scaling of the rebound angle \(\theta_1' \sim (1+\alpha/\beta) \sqrt{2(d_2-x)\theta_1} - \theta_1\), the total restitution \(e \sim \beta - (\beta^2-\alpha^2)(d_2 - x) \theta_1/\beta\), and the vertical restitution \(e_z \sim -\beta + (\alpha+\beta)\sqrt{2(d_2-x)/\theta_1}\) for \(f(x)\) yields the distributions
  \begin{align}
    \label{eq:2dvoid:shallow-impacts:rebound-angle-distribution}
    P(\theta_1' \vert \theta_1) &\sim
    \begin{dcases}
      \frac{\beta^2 (\theta_1+\theta_1')}{ (\alpha+\beta)^2 \sqrt 3 d_2 \theta_1} \ln \frac{2 (\alpha+\beta)^2  \sqrt 3 d_2 \theta_1}{\beta^2(\theta_1 + \theta_1')^2} \,,  & 0 < \frac{\beta(\theta_1+\theta_1') }{(\alpha + \beta) \sqrt{ \sqrt 3  d_2  \theta_1}} < 2 \,, \\
      0 \,, & \te{else,}
    \end{dcases} \\
    \label{eq:2dvoid:shallow-impacts:total-restitution-distribution}
    P(e \vert \theta_1) &\sim
    \begin{dcases}
      \frac{\beta}{(\beta^2-\alpha^2) \sqrt 3 d_2 \theta_1} \ln \frac{(\beta^2-\alpha^2) \sqrt 3 d_2 \theta_1}{\beta(\beta-e)} \,, & 0 < \frac{\beta(\beta-e)}{(\beta^2-\alpha^2) \sqrt 3 d_2 \theta_1} < 1 \,, \\
      0 \,, & \te{else,}
    \end{dcases} \\    \label{eq:2dvoid:shallow-impacts:vertical-restitution-distribution}
    P(e_z\vert \theta_1) &\sim
    \begin{dcases}
      \frac{(e_z + \beta) \theta_1^2}{(\alpha + \beta)^2 \sqrt 3 d_2 \theta_1} \ln \frac{2 (\alpha + \beta)^2  \sqrt 3 d_2 \theta_1}{(e_z+\beta)^2 \theta_1^2} \,, & 0 < \frac{(e_z+\beta)\theta_1}{(\alpha + \beta) \sqrt{ 2  \sqrt 3 d_2 \theta_1}} < 1 \,, \\
      0 \,, & \te{else,}
    \end{dcases}
  \end{align}
\end{widetext}
respectively.
The corresponding mean values
\begin{align}
  \label{eq:2dvoid:shallow-impacts:rebound-angle-mean}
  \overline{\theta_1'} &\sim (4/9)(1 + \alpha/\beta)\sqrt{2 \sqrt 3 d_2  \theta_1} - \theta_1 \,, \\
  \label{eq:2dvoid:shallow-impacts:total-restitution-mean}
  \overline e &\sim \beta - (\beta^2-\alpha^2) \sqrt 3  d_2  \theta_1/(8\beta) \,, \\
  \label{eq:2dvoid:shallow-impacts:vertical-restitution-mean}
  \overline {e_z} & -\beta + (4/9)(\alpha+\beta)\sqrt{2 \sqrt 3 d_2 / \theta_1} \,.
\end{align}
are of the form of the corresponding expressions for the two-dimensional bed in Eqs.~\eqref{eq:shallow-impacts:rebound-angle-mean}, \eqref{eq:shallow-impacts:total-restitution}, and \eqref{eq:shallow-impacts:vertical-restitution}, with a renormalized dimensionless bed grain size \(d_2\).

Although our analytical expressions rely on quite drastic simplifications, their agreement with the full model is good enough for computing qualitatively reliable predictions of the rebound statistics.
As an illustration, we compare the approximate relation for the rebound angle distribution, Eq.~\eqref{eq:2dvoid:shallow-impacts:rebound-angle-distribution}, with the numerical solution of the 3D model for three different grain-size ratios in Fig.~\ref{fig:rebound-angle-distribution}.

\subsection{\label{sec:exp-sim}Comparison with simulations and experiments}
We now test the predictions of the various versions of our model against experiments and computer simulations.
In summary, we find that the simplest two-dimensional approach, even without a second collision, suffices to fit the rebound averages for monodisperse granulates, \ie, as long as \(d_1/d_2 = 1\), whereas the dependence of these averages on \(d_1/d_2\) can only qualitatively be reproduced by the two-dimensional models, while quantitative predictions actually require some three-dimensional information about the bed packing.

\begin{figure*}
  \centering
  \includegraphics[width=\linewidth]{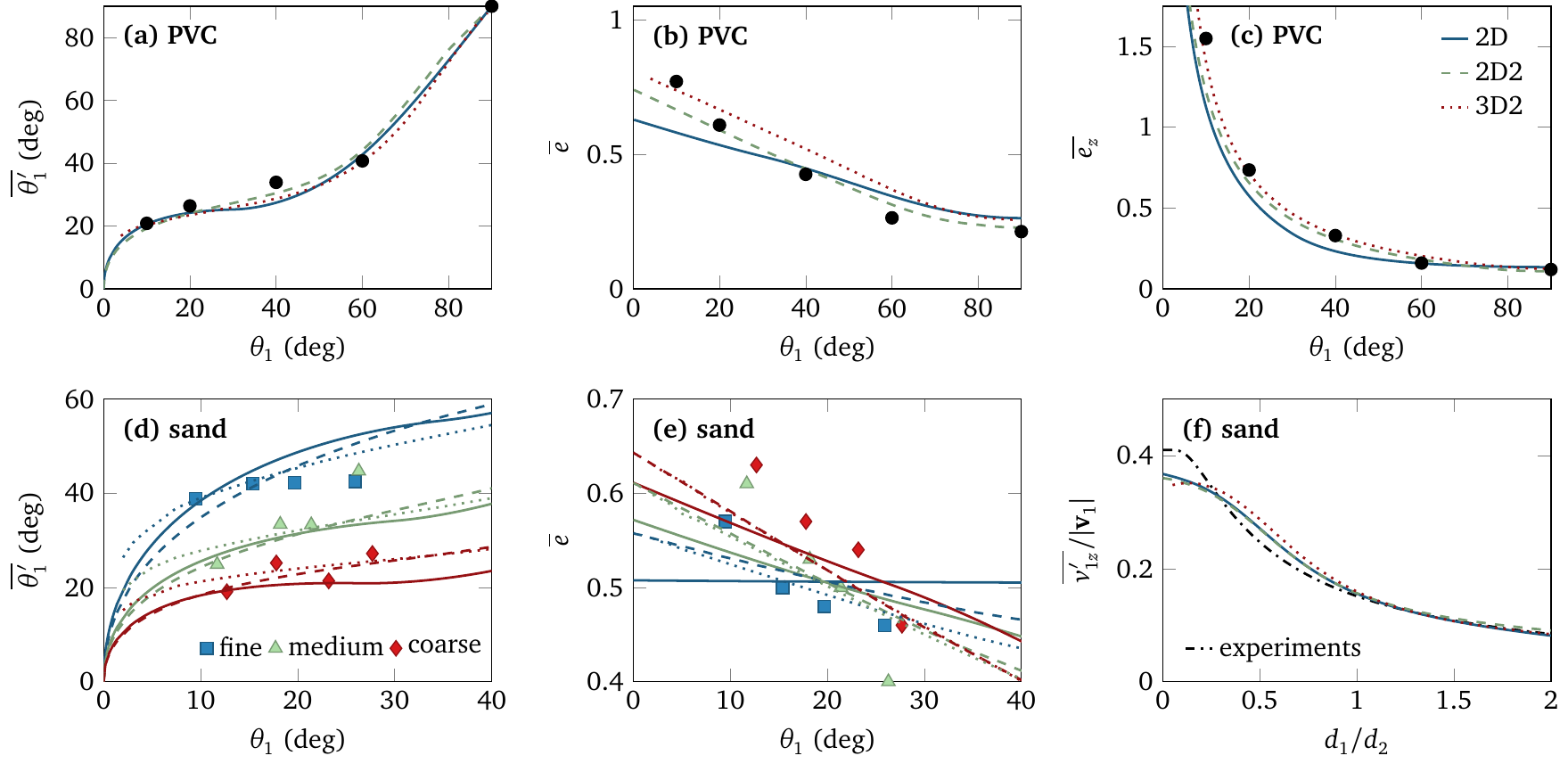}
  \caption{\label{fig:exp-sim:experiments}{Comparison with literature data.}
      [(a--c)] The two-dimensional model with (2D2) and without (2D) a second bed collision compared to the full three-dimensional version (3D2) and experimental data by Beladjine \etal~\cite{Beladjine2007} (dots) obtained for plastic PVC beads.
    Each model version is fitted to the data for the mean rebound angle \(\overline{ \theta_1'}\), the total restitution \(\overline e\), and the vertical restitution \(\overline {e_z}\) as a function of the impact angle \(\theta_1\) using the microscopic restitution coefficients \(\epsilon\) and \(\nu\) as global fit parameters.
    [(d, e)] Similar model fits to the wind-tunnel data by Willetts and Rice~\cite{Willetts1989} (symbols), who discriminated among fine, medium, and coarse grain fractions of the used sand sample to analyze their measurements.
    (f) Also the empirical relation \(\overline{v_{1z}'}/\abs{\vv v_1} = 0.41(1-10^{-0.2 d_2/d_1}) \) for grain-size dependence of the mean vertical rebound velocity for fixed impact angle \(\theta_1 = \SI{14}{\degree}\), proposed by Ellwood \etal~\cite{Ellwood1975} to fit their collision experiments with natural sand, is qualitatively well reproduced.
    Values of the microscopic restitution coefficients \(\epsilon\) and \(\nu\) for all shown fits are given in Table~\ref{tab:exp-sim:fit-parameters}.}
\end{figure*}

We start with the collision experiments by Beladjine \etal~\cite{Beladjine2007}, who shot plastic beads into a bed of similar beads to obtain the mean rebound angle \(\overline{ \theta_1'}\), the total restitution \(\overline e\), and the vertical restitution \(\overline {e_z}\) as a function of the impact angle.
As shown in Figs.~\ref{fig:exp-sim:experiments}(a)--\ref{fig:exp-sim:experiments}(c), the numerically evaluated two and three-dimensional model versions compare well with these data.
For each version, the values of the (effective) microscopic restitution coefficients \(\epsilon\) and \(\nu\), which are used as global fit parameters, are listed in Table~\ref{tab:exp-sim:fit-parameters}.

The dependence of the splash properties on the grain-size ratio was addressed in only very few experimental studies so far.
Willetts and Rice~\cite{Willetts1989}, for instance, used dune sand that is characterized by a unimodal grain-size distribution ranging from about 150 to \SI{600}{\micro m}, which they split into three fractions---fine, medium, and coarse---to investigate the influence of the size of the impactor on the rebound.
From their data we infer the grain-size ratios \(d_1/d_2 \approx 0.73\), \(1\), and \(1.4\), whereby we identified \(d_1\) with the mean diameter of the fine, medium, or coarse grain fraction and \(d_2\) with the overall mean.
The authors recorded the collision process in a wind tunnel during saltation, \ie, when the grains are driven by the wind.
Altering the bed inclination, they were able to tune the impact angle \(\theta_1\) of the fast hopping grains and thereby to investigate its influence on the mean rebound angle \(\overline{\theta_1'}\) and the total restitution \(\overline e\).
Again, we globally fit these data using \(\epsilon\) and \(\nu\) as free fit parameters.
The result, shown in Figs.~\ref{fig:exp-sim:experiments}(d) and \ref{fig:exp-sim:experiments}(e), reveals that the influence of the varying grain size is convincingly reproduced by both the two- and three-dimensional approach.
To improve the qualitative agreement between the data and the 3D model version, we here manually set the grain-size ratios to \(d_1/d_2 = 0.75\), \(1\), and \(1.25\), which corresponds to a slightly smaller polydispersity of the sand sample than expected form the measured grain-size distribution.

A few years before Willetts and Rice, Ellwood \etal~\cite{Ellwood1975} used sieved natural sand to measure the vertical rebound speed \(v_{1z}'\) for a fixed impact angle \(\theta_1 = \SI{14}{\degree}\) and various size ratios \(d_2/d_1\).
From their data, they extracted an empirical formula \(\overline{v_{1z}'}/\abs{\vv v_1} = 0.41(1-10^{-0.2 d_2/d_1}) \) that we compare Fig.~\ref{fig:exp-sim:experiments}(f) with the predictions of the two- and three-dimensional versions of our collision model.
Up to minor quantitative deviations in the limit of fine impactors, \(d_1/d_2<0.2\), all model versions are found to be in very good qualitative agreement with the empirical formula, which strongly supports our choice for the dependence of the restitution coefficients \(\alpha\) and \(\beta\) on the mass ratio \(d_1^3/d_2^3\) in Eqs.~\eqref{eq:restitution-coefficients} and \eqref{eq:effective-massratio}.

\begin{figure}
  \centering
  \includegraphics[width=\linewidth]{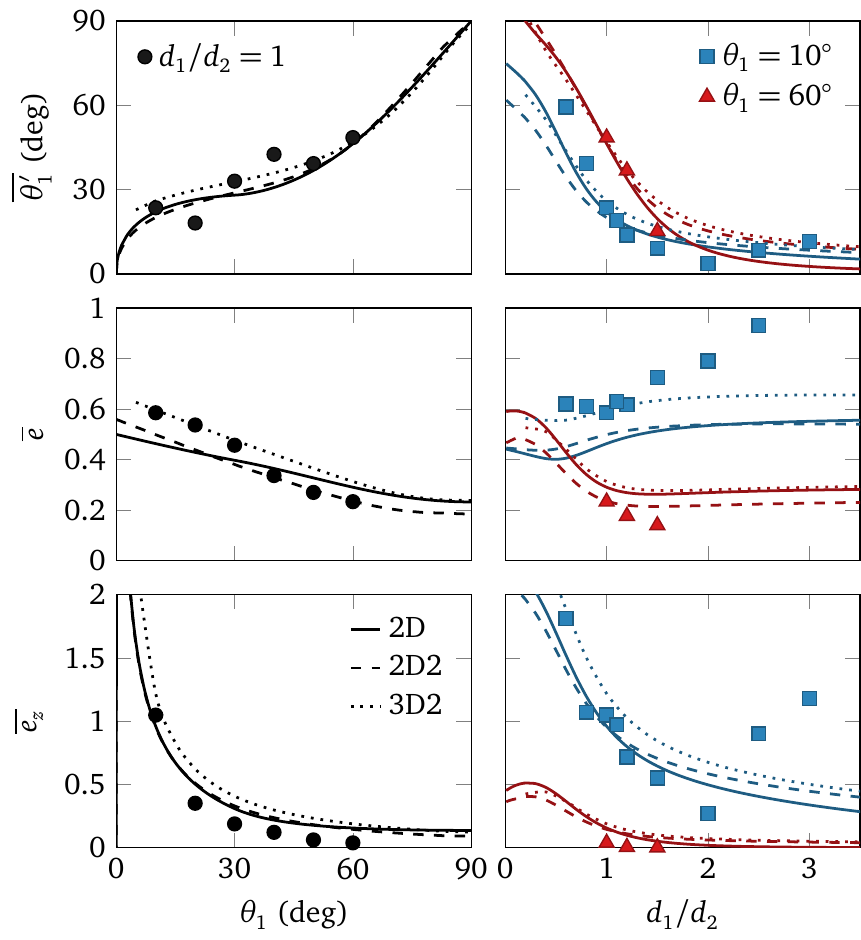}
  \caption{\label{fig:exp-sim:sim-luc-means}Comparison with computer simulations: averaged rebound characteristics as a function of the impact angle \(\theta_1\) and the grain-size ratio \(d_1/d_2\), as predicted by the 2D and 3D model versions and compared with computer simulations that describe the collision of an impacting bead with a 3D packing using a discrete-element method~\cite{Oger2005,Oger2013}.}
\end{figure}

\begin{figure}
  \centering
  \includegraphics[width=\linewidth]{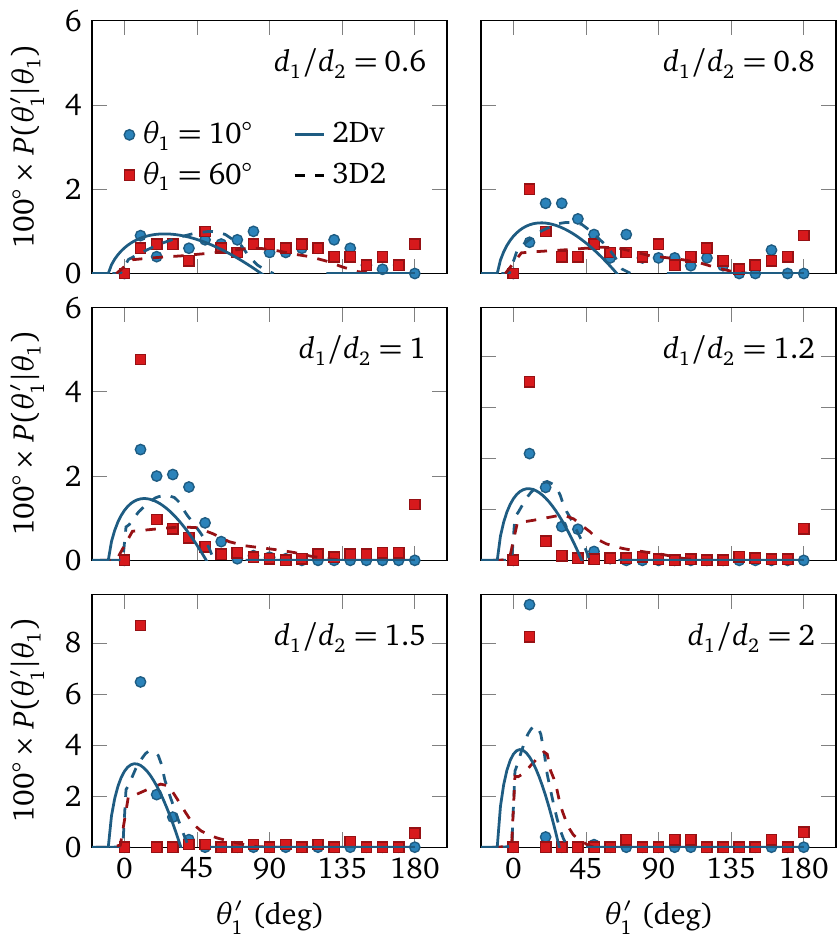}
  \caption{\label{fig:exp-sim:sim-luc-thdistr}Comparison with computer simulations: rebound angle distribution from the discrete-element simulations (symbols) compared to the three-dimensional collision model (3D2) of Sec.~\ref{sec:3d-collision-model} and the asymptotic scaling relation Eq.~\eqref{eq:2dvoid:shallow-impacts:rebound-distribution} for shallow impacts obtained from the effective two-dimensional model (2Dv) with a uniform void-size distribution.}
\end{figure}

As all the currently available experimental data are limited to rather confined parameter ranges---in particular, laboratory studies on the influence of the grain-size ratio are still lacking---we also test our model predictions against computer simulations that allow us to freely tune these parameters.
Details about the discrete-element method that is employed to simulate the collision of an impacting bead with a three-dimensional packing of beads of varying size can be found in Refs.~\cite{Oger2005,Oger2013}.
The dissipative collision between two beads at contact are quantified in terms of a friction coefficient and a normal restitution coefficient, which we set to \(0.3\) and \(0.8\), respectively.
As we found the rebound properties to be independent (within the statistical error bars) of the impact speed \(\abs{\vv v_1}\), we average all observables over the used \(\abs{\vv v_1} = 20\), 30, and \SI{40}{m/s}.
Figure \ref{fig:exp-sim:sim-luc-means} shows the so obtained mean rebound angle \(\overline{\theta_1'}\), the total restitution \(\overline e\), and the vertical restitution \(\overline{e_z}\) as a function of the impact angle \(\theta_1\) and the impactor--bed grain-size ratio \(d_1/d_2\) and compares them with the corresponding predictions of the two- and three-dimensional version our collision model.
Again, we used the (effective) microscopic restitution coefficients \(\epsilon\) and \(\nu\) as global fit parameters for each model version.
Their values are listed in Table~\ref{tab:exp-sim:fit-parameters}.
Besides such average quantities, the simulations provide us with the full rebound statistics.
As an example, we compare in Fig.~\ref{fig:exp-sim:sim-luc-thdistr} the distribution of the rebound angle with the numerical solution of the three-dimensional model version with a second collision and the asymptotic relation for shallow impacts derived form the two-dimensional effective approach with a uniform void-size distribution, introduced in Sec.~\ref{sec:2d-void}.

\begin{table}
  \caption{\label{tab:exp-sim:fit-parameters}Values of the microscopic restitution coefficients \(\epsilon\)  and \(\nu\) employed to fit the collision experiments with plastic beads by Beladjine \etal~\cite{Beladjine2007}, with natural sand by Willetts and Rice~\cite{Willetts1989} and Ellwood \etal~\cite{Ellwood1975}, and our discrete-element simulations~\cite{Oger2005,Oger2013} by our  two-dimensional model with (2D2) and without (2D) a second bed collision and its three-dimensional extension with a second bed collision (3D2).}
  \begin{ruledtabular}
    \begin{tabular}{lldddd}
      & & \multicolumn{1}{c}{plastic~\cite{Beladjine2007}} & \multicolumn{1}{c}{sand~\cite{Willetts1989}} & \multicolumn{1}{c}{sand~\cite{Ellwood1975}} & \multicolumn{1}{c}{simulations} \\
      \colrule
      \multirow{2}{*}{2D}  & \(\epsilon\) &  0.70 &  0.94 &  0.46 &  0.67 \\
      & \(\nu\)      & -0.83 & -1.2  & -0.86 & -1.5  \\
      \multirow{2}{*}{2D2} & \(\epsilon\) &  0.58 &  0.84 &  0.53 &  0.53 \\
      & \(\nu\)      & -0.24 & -0.98 & -1.4  & -1.1  \\
      \multirow{2}{*}{3D2} & \(\epsilon\) &  0.58 &  0.86 &  0.68 &  0.79 \\
      & \(\nu\)      & -0.24 & -0.97 & -1.6 & -0.69 \\
    \end{tabular}
  \end{ruledtabular}
\end{table}

\section{\label{sec:bed-grain-ejection}Bed grain ejection}
We finally turn to the theoretical description of the actual splash, the ejection of bed grains by the impacting particle.
High-speed videos of the collision process reveal that bed grains are not directly knocked out of the assembly by the impactor, but that they rather leave from a relatively large area of the bed, shortly after the impinging grain already bounced off (see, \eg, [SI]).
This suggests a partial fluidization of the bed caused by the momentum and energy transfer through the initially quiescent grain packing.

Valance and Crassous~\cite{Valance2009} modeled this process theoretically by a numerical scheme that could be extended to binary mixtures.
They further proposed an alternative energy diffusion approach that is easier to analyze, which could also serve as a starting point for further studies.
As yet another approach, one could consider the force propagation along \emph{force chains} in a granular bed.
Thereby one can, for example, account for the pressure dip beneath the apex of a sand pile produced by depositing grains from a nozzle.
Its occurrence depends on the history of the granular packing (\ie, the preparation of the pile) and can be suppressed by strong disorder and intergrain friction~\cite{Goldenberg2005}.
This behavior can be linked to two different mathematical descriptions of the stress propagation in static granular media, corresponding to hyperbolic or elliptic differential equations. The former gives rise to a type of ray propagation along force chains and the second to ``diffusive'' stress fields similar to those in homogeneous elastica (see, \eg, the lecture notes by Bouchaud~\cite{Bouchaud2002} for an overview).
One can actually derive these macroscopic relations from simple force balances on the grain contact level.
However, this is only possible for a static packing under the influence of gravity, where the weight is transferred downwards from one grain layer to the other.
For momentum propagation the situation is far more complex:
The grains move and may therefore change the contact network; and one has to account for momentum changes of all collision partners.
A possible starting point might be the force chain splitting approach by Bouchaud and coworkers~\cite{Bouchaud2001,Bouchaud2002,Socolar2002}, who proposed simple rules to create scattering paths through a quenched random packing.
Originally interpreted as static force chains, the very same paths might tentatively be used to model momentum propagation, which is highly suggested by the experiments by Clark and colleagues, who analyzed the force distribution~\cite{Clark2012,Clark2015} and the flow field~\cite{Clark2016} in a granular bed hit by a large intruder.
Particularly striking are the videos of these experiments, see also Ref.~\cite{Schirber2012} for an example and a brief review of this work.
The quasi-static modeling approach by Bouchaud \etal~was criticized, \eg, by Wyart~\cite{Muller2015}, for its lack of floppy modes, which turn out to be essential for a proper understanding of the response of granular packings to weak forces, in particular, for the characteristic power law distribution of small forces in the network.
In our case, however, we may tentatively argue that the force exerted by the fast impacting grain is large and the tail of the weak contact forces is inconsequential for the splash, for which gravity has to be overcome.

The approach we want to pursue here is mainly inspired by the above described picture of the branching force chains.
We follow Ho \etal~\cite{Ho2012}, who estimated the velocity distribution of the ejected particles by mapping the cascade of collisions in the packing to a fragmentation process.
As they were interested in the generic shape of this distribution, they assumed that in each collision, the kinetic energy is equally distributed among two target grains without any losses.
The energy transferred to a particle at the end of a chain of \(k\) collisions is thus given by the fraction \((1/2)^k\) of the energy transferred to the bed.
The latter could be estimated by \((1-\overline e^2) E_1\), where \(\overline e\) denotes the mean total restitution, as introduced in the previous section, and \(E_1 = (m_2/2) \vv v_1^2\) is the kinetic energy of the impactor.
The collisions in the packing can easily be made dissipative by replacing the energy-splitting factor \(1/2\) by a smaller effective restitution coefficient, but this would not affect the structure of the final result.
Ho \etal\ argued that the following fragmentation process can be used to describe the collision cascade:
The energy is split into two fractions of equal size.
Next, one of the resulting fractions is selected randomly and again split into two.
Then, one of the three fractions is selected and split, and so on.
It is well known that the energy fractions created through such a procedure are Poisson distributed with a parameter \(\lambda\) determined by the total number of splitting events.
For the splash process, this number is given by the total input energy divided by the minimum ejection energy, \ie, \((1-\overline e^2) E_1/(m_2 g d_2)\).
For typical impact speeds (\(>10\sqrt{g d_1}\)) and not too small grain-size ratios (\(d_1/d_2 > 0.5\)), \(\lambda\) is large enough so the Poisson distribution can be approximated by a normal distribution.
This eventually yields a log-normal distribution
\begin{equation}
  \label{eq:fragmentation-energy-distribution}
  P(E_2 \vert E_1) = \frac{1}{\sqrt{2\pi} \sigma E_2} \exp \! \left[-\frac{(\ln E_2-\mu)^2}{2\sigma^2} \right]
\end{equation}
for the energy \(E_2\) of the ejected particles, where
\begin{equation}
  \label{eq:fragmentation:energy-sigma-mu}
  \begin{split}
    \sigma &= \sqrt \lambda \ln 2\,, \\ 
    \mu &= \ln [(1-\overline
    e^2)E_1] - \lambda \ln 2 \,, \\ 
    \lambda &= 2 \ln\left[\left(1-\overline e^2\right)E_1/E_{d_2}\right] \,,
  \end{split}
\end{equation}
and \(E_{d_2} \equiv m_2 g d_2\) is the minimum transferred energy for a bed particle to be counted as ejecta.
It was shown in Ref.~\cite{Ho2012} that Eq.~\eqref{eq:fragmentation-energy-distribution}, rewritten in terms of the ejection velocity \(\abs{\vv v_2'} = \sqrt{2 E_2/m_2}\), is in excellent agreement with simulations of a discrete collision model and even with wind-tunnel measurements of saltating particles.
Both the simulations and the experiments were performed with unimodal sand, \ie, \(d_1/d_2=1\), and with values for the parameter \(\lambda\) varying between \(8\) and \(17\).
  However, even for grain-size ratios below \(0.5\), corresponding to \(\lambda\) on the order of 1,  the minor quantitative errors incurred by the log-normal approximation would not appreciably affect the following qualitative predictions.

\begin{figure}
  \centering
 \includegraphics[width=\linewidth]{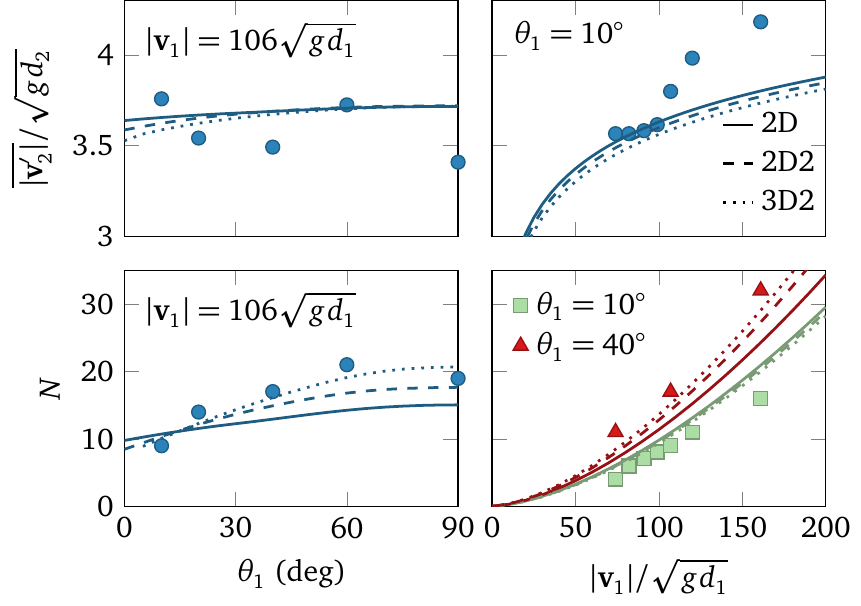}
 \caption{\label{fig:exp-sim:exp-beladjine2007-ejecta}Comparison with experimental literature data: The measured key characteristics of the ejected particles in the splash caused by an impacting grain can be reproduced by the fragmentation model by Ho \etal~\cite{Ho2012} if combined with our collision model from Sec.~\ref{sec:rebound}.
   To compare the mean ejection velocity \(\overline{\abs{\vv v_2'}}\), Eq.~\eqref{eq:ejection-velocity}, and the number \(N\) of ejected grains, Eq.~\eqref{eq:number-ejected-particles}, over a wide range of impact angles \(\theta_1\) and impact speeds \(\abs{ \vv v_1}\) with the experiments by Beladjine \etal~\cite{Beladjine2007} (symbols), we inserted the values of the microscopic restitution coefficients \(\epsilon\) and \(\nu\) listed in Table~\ref{tab:exp-sim:fit-parameters}.
   The fraction \(\gamma\) of the impact energy lost in the bed was used as fit parameter, which yields \(\gamma = 0.049\), \(0.055\), and \(0.062\) for the 2D, 2D2, and 3D2 versions of the collision model, respectively.}
\end{figure}

We here test Eq.~\eqref{eq:fragmentation-energy-distribution}, together with the Eq.~\eqref{eq:fragmentation:energy-sigma-mu}, against the laboratory data by Beladjine \etal~\cite{Beladjine2007}.
From these relations, we can calculate the mean ejection velocity
\begin{equation}
  \label{eq:ejection-velocity}
  \begin{split}
    \overline{\abs{\vv v_2'}} &= \int_{E_{d_2}}^\infty \hspace{-1.1em}
    \df E_2 \sqrt{2 E_2/m_2} P(E_2|E_1) \Bigg/ \int_{E_{d_2}}^\infty
    \hspace{-1.1em} \df E_2 P(E_2|E_1) \, \\
    &= \frac{\erfc[(\ln E_{d_2} - \mu-\sigma^2/2)/(\sqrt 2 \sigma)] }{ \erfc[(\ln E_{d_2} - \mu)/(\sqrt 2 )] } \sqrt 2 \te e^{\mu/2+\sigma^2/8} \,,
  \end{split}
\end{equation}
from the reduced ensemble of mobilized grains with energy \(E_2 \geq E_{d_2}\).
We thereby obtain its dependence on the impact angle and the impact speed, which we compare with the experimental data in Fig.~\ref{fig:exp-sim:exp-beladjine2007-ejecta}.
Recalling that the shown curves are \emph{not fitted} to the data, as there is no free parameter left in Eq.~\eqref{eq:ejection-velocity}, we find the agreement very satisfactory---only for the highest impact speeds the theory seems to underestimate the measured ejection velocities.
Moreover, the plot reveals that \(\overline{\abs{\vv v_2'}}\) depends only very weakly on the choice of the model version used to compute the total restitution \(\overline e\), which underscores that the fragmentation approach indeed robustly captures the underlying physics.

The fragmentation model as such does not provide us with a prediction for the total number \(N\) of ejected particles, which is an important coarse-grained measure of the splash function and frequently used in transport models to parametrize the splash.
But, following the same lines as for the mean ejection speed, we may combine it with the rebound properties obtained in Sec.~\ref{sec:rebound} to estimate \(N\) and its dependence on impact speed, impact angle, and grain-size ratio.
Subsuming the energy losses in the bed packing into the numerical prefactor \(\gamma\), the energy that goes into the mobilized (not necessarily ejected) grains can be written as \(\gamma ( 1 - \overline e^2) E_1\).
Divided by the average energy \(\overline{ E_2}\) of one mobilized grain it yields the total number of mobilized grains, from which we obtain the number
\begin{equation}
  \label{eq:number-ejected-particles}
  \begin{split}
    N &\approx \gamma \frac{\left(1-\overline
        e^2\right)E_1}{\overline{E_2}} \int_{E_{d_2}}^\infty
    \hspace{-1.1em} \df E_2 P(E_2|E_1) \\
    &= \gamma
    \frac{\left(1-\overline e^2\right)E_1}{2\overline{E_2}}
    \erfc\left( \frac{\ln E_{d_2}-\mu}{\sqrt2 \sigma} \right)\,,
  \end{split}
\end{equation}
of ejected grains.
In the second step, we used the log-normal energy distribution, Eq.~\eqref{eq:fragmentation-energy-distribution}, from which we also estimate the average energy
\begin{equation}
  \begin{split}
    \overline{ E_2} &= \te e^{\mu+\sigma^2/2} \\
    &= E_{d_2} \left[
      (1-\overline e^2)E_1/E_{d_2} \right]^{1-(2-\ln 2)\ln 2} \,,
  \end{split}
\end{equation}
of a mobilized grain.
The small exponent \(1-(2-\ln 2)\ln 2 \approx 0.1\) implies that the mean ejection energy is actually on the same order as its minimum value \(E_{d_2}\).

In Ref.~\cite{Valance2009}, \(\gamma\) was estimated from an energy-diffusion model, which allows to trace back the large energy losses in the bed to a simple geometrical effect: only a very small fraction of the downward propagating impact energy is scattered back towards the bed surface by collisions in the bed packing.
This means that, even for elastically colliding bed grains, typical values of \(\gamma\) are expected to be on the order of a few percent.
In fact, the values predicted by the diffusion model without energy dissipation varied between 0.05 and 0.2 for small to large impact velocities.
In the lower panels of Fig.~\ref{fig:exp-sim:exp-beladjine2007-ejecta}, Eq.~\eqref{eq:number-ejected-particles} is tested against the plastic-bead experiments by Beladjine \etal, where the total restitution \(\overline e\) is taken from the collision models presented in Sec.~\ref{sec:rebound} and \(\gamma\) is used as fit parameter.
Depending on the model version, \(\gamma\) varies between 0.052 and 0.071, which is on the same order as estimated in Ref.~\cite{Valance2009} for moderate impact speeds.

\section{\label{sec:analytical-splash}A ready-to-use analytical  splash parametrization}
In this section, we propose an exemplary list of relations that can be used to compute all needed splash properties for shallow impacts, as it is, for example, required in coarse-grained simulations of aeolian particle transport.
From Sec.~\ref{sec:2d-void} and Figs.~\ref{fig:rebound-angle-distribution} and \ref{fig:exp-sim:sim-luc-thdistr}, we conclude that the bumpiness of the bed surface yields only minute fluctuations in the total restitution coefficient, whereas the rebound angle varies significantly and may therefore be held responsible for the stochastic nature of the particle trajectories.
We therefore fix the total restitution coefficient to its mean \(\overline e\), as given by the shallow-impact asymptotics in Eq.~\eqref{eq:shallow-impacts:total-restitution}, and use Eq.~\eqref{eq:2dvoid:shallow-impacts:rebound-angle-distribution} for the rebound-angle distribution \(P(\theta_1' \vert \theta_1)\).
For both expressions, we employ the same values for the microscopic restitution coefficients \(\epsilon\) and \(\nu\), taken from, say, the fits for the 2D model version as given in Table~\ref{tab:exp-sim:fit-parameters}.
(To unify the various modeling approaches, the bed grain size \(d_2\) Eq.~\eqref{eq:2dvoid:shallow-impacts:rebound-angle-distribution} may be replaced by \(9d_2/(4\sqrt{3})\), so it yields the same asymptotic relation for the mean rebound angle as the 2D model version.
This has only marginal consequences for the shape of the rebound-angle distribution.)

An interesting measure that is typically required as input for coarse-grained particle transport simulations~\cite{Anderson1991,Andreotti2004,Kok2009} is the rebound probability \(P_\te{reb}(\vv v_1)\).
Within our approach, it can be defined as the probability that the vertical rebound speed of the impacting grain is larger than \(\sqrt{2 g d_1}\).
With Eqs.~\eqref{eq:shallow-impacts:total-restitution} and \eqref{eq:2dvoid:shallow-impacts:rebound-angle-distribution} it becomes
\begin{equation}
  \label{eq:rebound-probability}
  \begin{split}
    P_\te{reb}(\vv v_1) &= \int \!\! \df \theta_1' P(\theta_1' \vert
    \theta_1) \, \Theta\!\left(\overline e \abs{\vv v_1} \sin \theta_1'
      - \sqrt{2 g d_1} \right) \\
    &= 1-\frac{1+\ln \xi}{\xi} \,,
  \end{split}
\end{equation}
with
\begin{equation}
\xi \equiv  \frac{ 2\sqrt 2(\alpha+\beta)^2 d_2 \theta_1}{\beta^2(\theta_1 + \sqrt{2gd_1}/\abs{\vv v_1})^2} \,,
\end{equation}
which is indeed in good qualitative agreement with the parametrization \(P_\te{reb}(\vv v_1) \propto 1 - \exp\left( -\abs{\vv v_1}/ v_\te c \right) \) proposed by Anderson and Haff~\cite{Anderson1991} based on their grain-scale computer simulations of the splash process.
However, the magnitude of \(P_\te{reb}\) predicted by Eq.~\eqref{eq:rebound-probability} is slightly smaller than expected, because our single-collision approximation for \(P(\theta_1' \vert \theta_1)\), Eq.~\eqref{eq:2dvoid:shallow-impacts:rebound-angle-distribution}, yields a significant fraction of negative rebound angles.
Comparing this analytical estimate with the numerical solution that accounts for a second bed collision (\ie, version 3D2) in Fig.~\ref{fig:exp-sim:sim-luc-thdistr}, reveals that one should shift the analytical rebound-angle distribution to strictly positive values, namely as \(\theta_1' \mapsto \theta_1' + \theta_1\), to account for this effect in a simple way.
The full splash parametrization is completed by Eq.~\eqref{eq:number-ejected-particles}, with \(\gamma=0.06\), and Eq.~\eqref{eq:fragmentation-energy-distribution}, which determine the number of ejected particles and their velocity distribution, respectively.

\section{\label{sec:summary}Summary}
In this contribution, we aimed at a manageable parametrization of the splash function, which rests on basic physical principles, like momentum conservation and energy dissipation through inelastic pair collisions of the grains.
To this end, we started from a geometrical description of the collision of a spherical grain with a regular granular packing.
We introduced semi-phenomenological expressions for the gain-size-dependent normal and tangential restitution coefficient for such a grain--bed collision.
These two coefficients depend on two microscopic restitution coefficients \(\epsilon\) and \(\nu\), for the normal and tangential velocity losses during the inelastic grain--grain collisions, which serve as fit parameters in the model.
This approach eventually yields the rebound velocity of an impacting grain as a function of the impact velocity and the impactor--bed grain-size ratio.
We completed our parametrization of the splash function by combining this framework for the rebound with the energy-splitting model by Ho \etal~\cite{Ho2012}, which predicts how the impact energy is distributed among the bed grains.
It thereby gives access to the velocity distribution of the ejected bed grains and allows us to estimate how their total number scales with impact angle, impact speed, and grain-size ratio.

We have shown that the proposed two- and three-dimensional versions of our collision model yield very similar predictions for typical observables of interest, like the mean rebound angle, the mean total, and the mean vertical restitution.
In general, we found that each model version can be convincingly fitted to various experiments and computer simulations, if we use \(\epsilon\) and \(\nu\) as free fit parameters.
Excellent agreement is obtained, in particular, for the two-dimensional model that accounts for secondary collisions with the bed. 
This is an important observation as this model version is simple and computationally relatively cheap, which makes it suitable for practical applications.
Moreover, the two-dimensional approach allows for analytical asymptotic relations for shallow impacts, as shown in Sec.~\ref{sec:2d:shallow-impacts}.
This limit is of particular relevance, because the trajectories of wind-blown hopping grains are characterized by very small impact angles on the order of \SI{10}{\degree}~\cite{Bagnold1941}.
Hence, the simple asymptotics might be used, for instance, in coarse-grained aeolian transport simulations that cannot afford to resolve the granular structure of the sand bed.
The three- and two-dimensional models yield almost the same dependence on the impact angle and the grain size ratio for the analyzed averages.
Only their distributions can differ qualitatively, the three-dimensional approach yielding smoother shapes for shallow impacts, as illustrated in Fig.~\ref{fig:rebound-angle-distribution}.
We showed that this shortcoming of the two-dimensional models can be overcome by an extension with a uniform distribution of void spaces between neighboring surface grains.
Thereby good agreement with the full three-dimensional model and with our discrete-element simulations could be achieved, as shown for the rebound-angle distribution in Fig.~\ref{fig:exp-sim:sim-luc-thdistr}.
Combined with the energy-fragmentation model by Ho \etal~\cite{Ho2012} for the statistics of the ejected bed particles, the simple two-dimensional impact model yields a ready-to-use parametrization for the splash.
It therefore provides an excellent starting point for modeling aeolian structure formation.
  This, however, requires some extensions of our parametrization, including the drag and lift forces due to the driving turbulent flow.
  Moreover, the inclusion of additional model ingredients, like the disaggregation of dust agglomerates due to collisions~\cite{Sullivan2008} or cohesive, hydrodynamic, and electrostatic interactions~\cite{Kok2012,Merrison2012}, could give rise to a much richer phenomenology. 
  They are of particular relevance for understanding exterrestrial granular structures, as observed on Mars~\cite{Sullivan2008} or, most recently, on a Jupiter comet~\cite{Thomas2015}.
  Dedicated theoretical approaches and experimental work~\cite{Rasmussen2015}, might help to extend our model to such phenomena in the future.

\begin{acknowledgments}
This research was supported by a Grant from the GIF, the German-Israeli Foundation for Scientific Research and Development.
  We also acknowledge the hospitality of the KITP in Santa Barbara and the MPI-PKS in Dresden, where this work was started, and financial support by the National Science Foundation under Grant No.\ NSF PHY-1125915, the MPI-PKS Visitors Program, and the German Academic Exchange Service (DAAD) through a Kurzstipendium (for M.L.)
  and the RISE program (for K.D.).
  M.L.\ thanks Maik We{\ss}ling for fruitful discussions during the early stages of this project.
\end{acknowledgments}

\appendix

\section{\label{sec:inel-binary-coll}Inelastic binary collisions}
We briefly outline the usual parametrization of an inelastic collision of two spheres in terms of the normal and tangential restitution coefficients \(\epsilon\) and \(\nu\), which account for dissipation of kinetic energy during the grain contact.
The energy loss by relative motion in normal direction originates from grain deformations, and the tangential loss characterizes the reduction of the relative velocity of the grain surfaces at the contact point due to friction.
The exact value of \(\nu\), which characterizes the tangential slip on particle contact, is hard to estimate, and we might, for simplicity, assume that the colliding spheres roll past each other, corresponding to \(\nu=0\).
However, comparing the model predictions obtained with experimental data in Sec.~\ref{sec:exp-sim}, we find that \(\nu\) has to be negative to fit the data, which means that the relative surface velocity (or the spin of the impactor) has formally to be reversed.
Exact results for the normal restitution \(\epsilon\) of perfect spheres are reviewed, \eg, in Ref.~\cite{Brilliantov2010}.
For viscoelastic Hertzian beads, one obtains that \(\epsilon\) decreases with the impact speed and the size of the colliding grains.
Corresponding marginal quantitative corrections to our discussion would not change the overall qualitative picture.

The surface velocities of the colliding grains are determined by the relative velocity of the centers of two colliding spheres and their rotational velocities.
The full calculation can be found in classical textbooks (see, \eg, the book by Brilliantov and P{\"o}schel~\cite{Brilliantov2010}), so we only give the result for the velocity \(\vv v_1'\) of the first grain after the collision,
\begin{equation}
  \label{eq:binary-collision-v1}
  \begin{split}
    \vv v_1' = \vv v_1 &- \frac{M}{m_1}(1+\epsilon) (\hat{\vv n} \cdot  \vv v_{12}) \hat{\vv n} \\
    &- \frac{M}{m_1} \frac{1-\nu}{1+q} (\mathds 1 - \hat{\vv n}\hat{\vv n}) \cdot \vv v_{12} \\
    &+ \frac{1}{2} \frac{M}{m_1} \frac{1-\nu}{1+q}  \hat{\vv n} \times (d_1 \vv \omega_1 + d_2 \vv \omega_2) \,.
  \end{split}
\end{equation}
Here \(d_{1,2}\) are the diameters of the two spherical grains, \(m_{1,2}\) their masses, and \(\vv v_{1,2}\) their velocities before the collision, which define \(\vv v_{12} \equiv \vv v_1 - \vv v_2\).
The unit vector \(\hat{\vv n}\) is parallel to the line that connects the centers of the spheres at contact.
The effective mass is \(M \equiv m_1 m_2/(m_1 + m_2) \) and the parameter \(q \equiv (M/4)(d_1^2/I_1 + d_2^2/I_2)\) depends on the moments of inertia \(I_{1,2}\) of the two grains.
For spheres, \(I_{1,2} = m_{1,2} d_{1,2}^2/10\) and thus \(q=5/2\).

Assuming that the colliding grains do not rotate and that the second grain is at rest before the collision, \(\omega_{1,2} = 0\) and \(\vv v_2 = 0\), as it is the case when an impactor hits the granular packing, Eq.~\eqref{eq:binary-collision-v1} reduces to
\begin{equation}
  \label{eq:binary-collision-v1-packing}
  \begin{split}
    \vv v_1' = &\left[1 - \frac{M}{m_1}(1+\epsilon) \right] \hat{\vv n} \hat{\vv n} \cdot \vv v_1 \\
    & \qquad + \left[ 1 - \frac{M}{m_1} \frac{1-\nu}{1+q} \right] (\mathds 1 - \hat{\vv n}\hat{\vv n}) \cdot \vv v_1 \,.
  \end{split}
\end{equation}
For grains of similar size, \(d_1 \approx d_2\), \(M \approx m_1/2\), and thus
\begin{equation}
  \label{eq:binary-collision-v1-packing-monodisperse}
  \vv v_1' \approx \left[ \frac{1-\epsilon}{2} \hat{\vv n} \hat{\vv n} + \frac{1+2q + \nu}{2+2q} (\mathds 1 - \hat{\vv n}\hat{\vv n})  \right] \cdot \vv v_1 \,.
\end{equation}
For small impactors, \(m_1 \ll m_2\), we may approximate \(M \sim m_1\), which yields
\begin{equation}
  \label{eq:binary-collision-v1-packing-small-impactor}
  \vv v_1' \sim \left[ \epsilon \hat{\vv n} \hat{\vv n} + \frac{\nu + q}{1+q} (\mathds 1 - \hat{\vv n}\hat{\vv n})  \right] \cdot \vv v_1 \,.
\end{equation}

\section{\label{sec:2d-steep-impacts}2D collision model: Steep impacts}
For steep impact, the impact position \(x_0\) is given by the second line of Eq.~\eqref{eq:integration-limits} and the integral in Eq.~\eqref{eq:rebound-distribution} simplifies to \( P(A \vert \vv v_1) \sim (1/d_2) \abs*{\df f^{-1}/\df A}\)
if the rebound condition \(-d_2 < 2 f^{-1}(A) - \sqrt{4-d_2^2} \cot \theta_1 < d_2\) is fulfilled and \(P(A \vert \vv v_1) \sim 0\) otherwise.
We start with the rebound angle \(f(x) = \theta_1'\), which we expand in the impact angle \(\theta_1\) around \(\theta_1 =\pi/2\),
\begin{equation}
  \label{eq:steep-impacts:rebound-angle-vs-x}
  \begin{split}
    \theta_1' \sim & \tan^{-1}\left[ \frac{\alpha - (\alpha + \beta)
        x^2}{(\alpha+\beta)x\sqrt{1-x^2}} \right] - (\theta_1-\pi/2) \\
    + &\frac{\alpha \beta + \alpha^2+( \beta^2-\alpha^2)
        x^2}{2\alpha^2+2( \beta^2-\alpha^2) x^2}
      \sqrt{\frac{4-d_2^2}{1-x^2}}(\theta_1-\pi/2) \,.
  \end{split}
\end{equation}
For convenience, we here substituted the shifted impact position \(x + \sqrt{1-(d_2/2)^2} \cot \theta_1\) for the argument of \(f(x)\), so that \(x\) takes values between \(-d_2/2\) and \(d_2/2\).
Integrating over this impact interval, we obtain Eq.~\eqref{eq:steep-impacts:rebound-angle-mean} of the main text for the mean rebound angle \(\overline{\theta_1'}\).
Although Eq.~\eqref{eq:steep-impacts:rebound-angle-vs-x} cannot be solved for \(x\), as required for the rebound angle distribution a closer look at Eq.~\eqref{eq:steep-impacts:rebound-angle-vs-x} reveals that it can be approximated by its first-order \(x\)-expansion \(\theta_1' \sim \pi/2 - (1+\beta/\alpha)x - [1- (1+\beta/\alpha) \sqrt{1-(d_2/2)^2}](\theta_1-\pi/2)\).
This, in turn allows to (roughly) estimate its inverse \(f^{-1}(\theta_1')\) and thus the asymptotic distribution of the impact angle.
Within this crude approximation, the latter evaluates to a uniform distribution,
\begin{widetext}
  \begin{equation}
    \label{eq:steep-impacts:rebound-angle-distribution}
    P(\theta_1' \vert \theta_1) \sim
    \begin{dcases}
      \frac{\alpha}{(\alpha+\beta) d_2} \,, & -d_2 < \frac{2\alpha(\theta_1 + \theta_1'-\pi)}{\alpha+\beta} + \sqrt{4-d_2^2}(\theta_1-\pi/2) < d_2 \,, \\
      0 \,, & \te{else. }
    \end{dcases}
  \end{equation}
Note that the mean rebound angle \(\overline{\theta_1'} \sim \pi/2 - \left[ 1 - ( 1 + \beta/\alpha ) \sqrt{4-d_2^2} \right] (\theta_1 - \pi/2)\) obtained from this approximate distribution differs from the correct asymptotic scaling relation given in Eq.~\eqref{eq:steep-impacts:rebound-angle-mean}.

Following the same idea, we obtain the \(x\)-dependence 
\begin{equation}
  \label{eq:steep-impacts:total-restitution-vs-x}
  e \sim \sqrt{\alpha^2+(\beta^2-\alpha^2)x^2} - (\beta^2-\alpha^2) x \sqrt{ \frac{1-(d_2/2)^2}{\alpha^2+(\beta^2-\alpha^2)x^2}} (\theta_1 - \pi/2)
\end{equation}
of the total restitution coefficient.
Integrated over the impact position \(x\), it yields the result for \(\overline e\) given in Eq.~\eqref{eq:steep-impacts:total-restitution-mean} of the main text.
Again, the distribution of \(e\) can analytically only be estimated from an approximate form of Eq.~\eqref{eq:steep-impacts:total-restitution-vs-x}, \eg, from the parabola \(e \sim \alpha + (\alpha/2)(\beta^2/\alpha^2-1)[x^2 - x \sqrt{4-d_2^2} (\theta_1-\pi/2)] \), which yields
\begin{equation}
  \label{eq:steep-impacts:total-restitution-distribution}
  P(e \vert \theta_1) \sim
  \begin{dcases} 
    \frac{1}{d_2}
\sqrt{ \frac{2\alpha}{(\beta^2-\alpha^2)(e-\alpha) }}   \,, & 
    0 < \frac{8\alpha(e-\alpha)}{\beta^2-\alpha^2} < d_2^2 - \sqrt{4-d_2^2}(\theta_1-\pi/2) \,, \\
    \frac{1}{d_2} \sqrt{ \frac{\alpha/2}{(\beta^2-\alpha^2)(e-\alpha) }}  \,,  & -1 < \frac{8\alpha(e-\alpha)/(\beta^2-\alpha^2)-d_2^2}{\sqrt{4-d_2^2}(\theta_1-\pi/2)} < 1 \,, \\
    0 \,, & \te{else,}
  \end{dcases}
\end{equation}
up to linear order in \(\theta_1-\pi/2\).
The first line represents the impact range
\(-d_2/2 < x < \sqrt{1-(d_2/2)^2}(\theta_1-\pi/2)\) where the inverse of \(e(x)\) has two branches; the second line corresponds to the single-branch region  \(\sqrt{1-(d_2/2)^2}(\theta_1-\pi/2) < x < d_2/2\).

Finally, we consider the steep-impact limit for the vertical restitution coefficient
\begin{equation}
  \label{eq:steep-impacts:vertical-restitution-vs-x}
  e_z \sim \alpha - (\alpha+\beta) x^2 + (\alpha+\beta) x \left(\sqrt{4-d_2^2} - \sqrt{1-x^2}\right) (\theta_1 - \pi/2) \,.
\end{equation}
Integrating over the impact positions \(x\), we obtain its mean \(\overline e\), given in Eq.~\eqref{eq:steep-impacts:vertical-restitution-mean}.
Again, Eq.~\eqref{eq:steep-impacts:vertical-restitution-vs-x} can be approximated by its second-order \(x\)-expansion \(e_z \sim \alpha - (\alpha+\beta) x^2 + (\alpha+\beta) x ( \sqrt{4-d_2^2} - \sqrt{1-x^2} ) (\theta_1 - \pi/2)\), from which we derive the estimate
\begin{equation}
  \label{eq:steep-impacts:vertical-restitution-distribution}
  P(e_z \vert \theta_1) \sim
  \begin{dcases}
    \frac{1}{2 d_2 \sqrt{(\alpha+\beta)(e_z-\alpha)}} \,, &  \te {if }  - 1 < \frac{d_2^2/2 + 2(e_z-\alpha)/(\alpha+\beta)}{ \left( 1-\sqrt{4-d_2^2} \right) (\theta_1-\pi/2)} < 1 \,, \\ 
    \frac{1}{d_2 \sqrt{(\alpha+\beta)(e_z-\alpha)}} \,, &  \te {if } \frac{d_2^2/2 + 2(e_z-\alpha)/(\alpha+\beta)}{ \left( 1-\sqrt{4-d_2^2} \right) (\theta_1-\pi/2)} > 1 \te{ and } e_z < \alpha\,, \\
    0\,, & \te{else.}
  \end{dcases}
\end{equation}
for the distribution of \(e_z\).
\end{widetext}

\bibliography{sand}

\end{document}